# Dynamic temperature compensation for wavelength-stable entangled biphoton generation

Yuting Liu[1,2], Huibo Hong[1,2], Xiao Xiang[1,]*, Runai Quan[1], Tao Liu[1,2,3], Mingtao Cao[1,2], Shougang Zhang[1,2,3] and Ruifang Dong[1,2,3]**

[1] Key Laboratory of Time Reference and Applications, National Time Service Center, Chinese Academy of Sciences, Xi'an 710600, China

[2] School of Astronomy and Space Science, University of Chinese Academy of Sciences, Beijing 100049, China

[3] Hefei National Laboratory, Hefei 230088, China

*Corresponding author: xiangxiao@ntsc.ac.cn; ** corresponding author: dongruifang@ntsc.ac.cn

A dynamic temperature compensation method is presented to stabilize the wavelength of the entangled biphoton source, which is generated via the spontaneous parametric down-conversion based on a MgO: PPLN waveguide. Utilizing the dispersive Fourier transformation technique combined with a digital proportional-integral-differential algorithm, the small amount of wavelength variation can be instantly identified and then compensated with active temperature correction. The long-term wavelength stability, assessed though Allan deviation, shows nearly a hundredfold enhancement, reaching $2.00 \times 10^{-7}$ at the averaging time of 10000 s. It offers a simple, ready-to-use solution for precise wavelength control in quantum information processing.

## Introduction

Entangled biphoton sources generated via the process of spontaneous parametric down-conversion (SPDC)[1-4] have played an important role in areas of quantum information technology[5, 6]. Leveraging the energy-time entanglement, these biphotons have showcased their remarkable utility across a diverse range of applications, ranging from quantum clock synchronization[7-11], quantum spectroscopy[12-14], quantum optical coherence tomography[15, 16], to quantum temporal ghost imaging[17, 18], etc. In these applications, precise control and stabilization of the photon wavelength is key to optimizing performance. For example, fine-grained nonlocal dispersion compensation has been found dependent on the precise wavelength tuning[19] and selecting[20]. To guarantee high-fidelity quantum interference, it is imperative to perform precise adjustment to align the frequency spectra of photons emanating from distinct sources[21-23]. Furthermore, in the fiber-optic two-way quantum time transfer system[8], the wavelength stability of the utilized photons has been recognized as a pivotal factor for improving the long-term time synchronization performance.

Nowadays, periodically poled nonlinear mediums are commonly employed for the generation of biphotons. The wavelength characteristics of the down-converted photons are influenced by both the phase-matching condition of the nonlinear medium and the wavelength of the pump laser. When a monochromatic laser at a certain wavelength is used as the pump, the wavelength stability of the down-converted photons primarily depends on the phase-matching condition of the nonlinear medium[24, 25]. Typically, temperature control is employed to achieve the desired phase-matching condition. This is accomplished by utilizing a thermo-electric cooler (TEC) and thermistor to tune and control the temperature of a well-designed package housing the nonlinear medium. The temperature of the package directly influences the working temperature of the nonlinear medium, which in turn determines the wavelength of the generated photons. However, due to the inevitable thermal gradient inside the nonlinear medium, the wavelength of the down-converted photons will experience a nontrivial drift despite the precise temperature control capability of the package [26, 27]. This drift becomes unacceptable for practical applications that rely on a stable wavelength operation. Taking our experimental results as an example, a 15.0 mm-long periodically polarized magnesium oxide doped lithium niobate waveguide (MgO: PPLN-WG) pumped by a 780 nm diode laser was utilized to generate the energy-time entangled biphotons. With an elaborate temperature control of the package housing the nonlinear medium at a precision of 10 mK, the wavelength uncertainty is expected to be reduced to less than 6 pm, as determined by calibration. However, during a measurement period of approximately 1.4 hours, we observed a significant wavelength drift of 342 pm, representing the temperature drift of the SPDC process with time. This drift would result in noticeable performance degradation for numerous applications. To enhance wavelength stability in entangled biphoton generation, an active temperature compensation method is needed. However, to the best of our knowledge, no such method has been reported thus far.

In this letter, a dynamic temperature compensation method is presented to actively stabilize the wavelength of the entangled biphoton source generated from the MgO: PPLN-WG. Utilizing the dispersive Fourier transformation (DFT) technique, the wavelength information of the generated photons is instantly recovered from the time-correlated single photon counting (TCSPC) histogram. By integrating a digital proportional-integral-differential (PID) algorithm, the dynamic temperature compensation can be further applied to stabilize the phase-matching condition of the waveguide. By this way, the long-term wavelength drift of the down-converted photons is significantly suppressed. The wavelength stability is evaluated in terms of Allan deviation (ADEV), which reaches $2.00 \times 10^{-7}$ over average time 10000 s. This represents a nearly hundredfold improvement compared to the stability obtained with the conventional static temperature control on the waveguide. The limiting factor to this improvement is primarily the wavelength stability of the pump laser in SPDC. This result presents a simple but efficient way for wavelength-stabilizing the entangled biphoton source generated from a temperature-controlled nonlinear medium. This achievement will greatly enhance the performance of numerical quantum information processing applications that rely on wavelength-stable biphoton sources.

## Methods

Given the unavoidable thermal gradient, the temperature distribution along the direction of the nonlinear medium becomes inconsistent. Consequently, under the condition that the nonlinear medium is TEC-controlled at $T_0$, the presence of variation in ambient temperature introduces a small but randomly varying temperature term over time, denoted as $\delta T$[26, 27]. Therefore, the actual temperature experienced by the SPDC process can be expressed as

$$T_{SPDC} = T_0 + \delta T. \tag{1}$$

Correspondingly, the actual center frequency of the SPDC generated photons from the nonlinear medium can be written as a temperature-dependent form:

$$\omega(T_{SPDC}) = \omega(T_0 + \delta T). \tag{2}$$

Approximating the function of $\omega(T)$ in Eq. (2) with a first-order Taylor expansion around $T_0$, one obtain

$$\omega(T_{SPDC}) \approx \omega(T_0) + \omega'\delta T. \tag{3}$$

Here $\omega'$ denotes the temperature-dependent frequency tuning coefficient, $\omega'\delta T$ represents the center frequency drift due to temperature variations during the SPDC process. To suppress such center frequency drift, we propose a dynamic temperature compensation method that integrates the DFT technique with digital PID algorithm. The underlying principles is outlined as follows.

The schematic diagram for resolving the spectral property of the idler photons using the DFT technique is shown in Fig. 1 (a). Out of the entangled biphoton source, the signal and idler photons are spatially departed. The idler photons are passed through a dispersive element with a dispersion parameter of $D$, while the signal photons are utilized as the reference. The measured coincidence distribution versus photon arrival time difference between the signal and idler photons (addressed as $\tau$) can be described as[14]

$$G^{(2)}(\tau) \propto exp\left(-\frac{(\tau-\tau_0)^2}{2\Delta^2}\right), \tag{4}$$

with its temporal width ($\Delta$) and center ($\tau_0$) being given by

$$\Delta = \frac{|D|\sigma_i}{\sqrt{2}}, \tag{5}$$

$$\tau_0 = D\omega_i(T_{SPDC}) + \tau_s, \tag{6}$$

where $\sigma_i$ denotes the idler photons' spectral width. According to the wavelength-to-time mapping relation established by DFT, $\tau_0$ is linearly dependent on the center frequency of the idler photons ($\omega_i(T_{SPDC})$) in addition to an intrinsic system latency ($\tau_s$). By inserting Eq. (3) into Eq. (6), we obtain

$$\tau_0 \approx D\omega_i(T_0) + D\omega_i'\delta T + \tau_s. \tag{7}$$

By measuring the variation of $\tau_0$ as a function of the preset temperature of the nonlinear medium, the temperature-dependent frequency tuning coefficient can be pre-determined by $\omega_i' = \Delta\tau_0/D$. Thus, as long as $T_0$, $\tau_s$, $D$ and $\omega_i'$ are fixed or pre-determined, the real-time temperature variation ($\delta T$) experienced by the SPDC process can be obtained through tracking the center position variation of the coincidence histogram. By utilizing the measured center position variation of the coincidence histogram as the feedback input to the TEC controller, which is based on a digital PID algorithm, we can then efficiently compensate for the temperature variation.

The flowchart of the dynamic temperature compensation scheme is illustrated in Fig. 1(b). Before implementing the compensation, the value of $\tau_0$ is initialized by the first measurement run of the coincidence center, denoted as $\tau_{0,ini} = D\omega_i(T_0) + \tau_s$. Based on the above-mentioned method, the temperature-dependent frequency tuning coefficient $\omega_i'$ is pre-determined. For the compensation implementation, the center position of the coincidence is continuously measured and compared with $\tau_{0,ini}$ to track the center position variation of the coincidence histogram. For the k-th measurement run ($k = 1, 2, 3 ...$), $\Delta\tau_{0,k} = \tau_{0,k} - \tau_{0,ini}$ denotes its associated center position variation, inferring a temperature variation ($\delta T_k$) experienced by the SPDC process.

Taking into account the limited measurement accuracy of the coincidence center, a threshold is preset to avoid ineffective feedback actions. Temperature compensation is only performed when the measured center position variation exceeds a preset threshold, i.e., $|\Delta\tau_{0,k}| \geq \tau_{th}$. In this case, the corresponding temperature variation can be determined by $\delta T_k = (\tau_{0,k} - \tau_{ini})/D\omega_i'$. To compensate for $\delta T_k$, it is subsequently processed using a digital PID algorithm composed of parameters $K_p$, $K_i$ and $K_d$ to generate the error signal ($\varepsilon_k$), which can be represented by

$$\varepsilon_k = K_p\delta T_k + K_i\sum_{j=0}^{k}\delta T_j + K_d(\delta T_k - \delta T_{k-1}). \tag{8}$$

Otherwise, if $|\Delta\tau_{0,k}| < \tau_{th}$, $\varepsilon_k = 0$. By accumulating the error signal and applying it to the TEC together with the pre-controlled temperature $T_0$, we can actively stabilize the phase matching temperature of the non-linear medium, thereby ensuring wavelength-stable entangled biphoton generation.

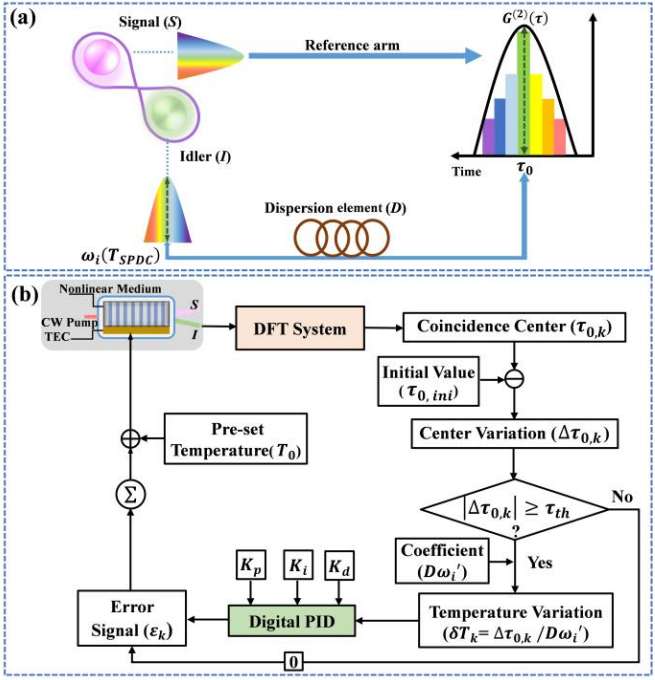

Fig.1. (a) Schematic diagram for spectral measurement of the single photons based on DFT. (b) Flowchart for stabilizing the frequency (wavelength) of the entangled biphoton source based on the dynamic temperature compensation. DFT System, the dispersive Fourier transformation system; TEC, thermo-electric cooler; PID, proportional-integral-derivative.

### Experiment

The experimental setup of dynamic temperature compensation for wavelength-stable entangled biphoton generation is presented in Fig. 2. The utilized nonlinear medium is a 15.0 mm-long, type-II MgO: PPLN-WG (HC Photonics Corp.) with a poling period of ~8.0 μm, which was pumped by a 780 nm continuous-wave laser (Thorlabs Inc., DBR-780PN). To implement the dynamic temperature compensation, a small portion (10 %) of the generated biphoton source was departed. The partitioned biphotons were then directed through a fiber polarizing beam splitter (FPBS) to spatially separate the orthogonally polarized signal and idler photons. Subsequently, the signal photons were detected directly using a superconducting nanowire single photon detector (SNSPD1, Photon Technology Ltd.), while their corresponding idler photons were detected by another superconducting nanowire single photon detector (SNSPD2) after passing through a fiber Bragg grating-based dispersion compensation module (FBG, Proximion Inc., DCMCB-SP-020P1FA). The outputs of the two SNSPDs were subsequently connected to a time-correlated single-photon counting module (TCSPC, Swabian Instruments, Time Tagger Ultra) for coincidence measurement. By applying Gaussian fitting to each run of the coincidence measurement, the variation of the coincidence center during the measurement time was obtained. Then, the corresponding temperature variation can be identified for subsequent compensation purposes. Due to the wavelength-to-time mapping relation associated with the DFT technology, the variation in the central frequency of the idler photons was accordingly determined with the pre-calibrated temperature-dependent frequency tuning coefficient.

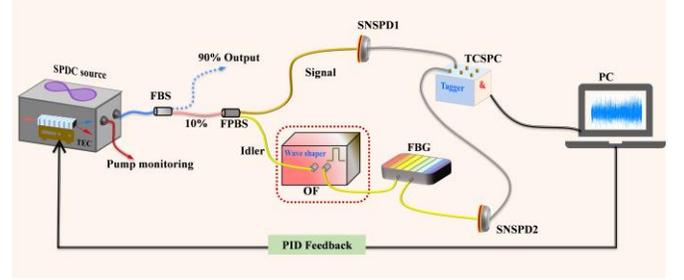

Fig. 2. Experimental setup with dynamic temperature compensation for wavelength-stable entangled biphoton generation. SPDC, spontaneous parametric down-conversion; TEC, thermo-electric cooler; FBS, 90/10 fiber beam splitter; FPBS, fiber polarization beam splitter; OF, tunable optical filter; FBG, fiber Bragg grating-based dispersion compensation module; SNSPD, superconductive nanowire single-photon detector; TCSPC, time-correlated single-photon counting module; PC, personal computer; PID, proportional-integral-derivative.

### Results and Discussion

As the mapping relation between the coincidence center and the photon wavelength center should be fixed in advance, the actual dispersion value of FBG needs to be calibrated accurately. By putting a tunable optical filter (OF, Finisar Wave-Shaper 1000B) at the front of the FBG, as shown in the virtual box of Fig. 2, we programmed the OF into a 0.80 nm bandpass filter and set its center wavelength ranging from 1559.79 nm to 1563.05 nm with a step of 0.80 nm. At each wavelength, the corresponding coincidence distributions were recorded. Through Gaussian fitting to these coincidence measurements, the temporal center of the coincidence as a function of the central wavelength of the OF was determined, as depicted by black dots in Fig. 3(a). By conducting a linear fitting to the results, represented by the red solid line, the group delay dispersion (GDD) of the FBG was determined, yielding a value of $(335.17 \pm 10.07)$ ps/nm around 1560 nm.

With the OF being set in the full-pass mode, by varying the working temperature of MgO: PPLN-WG from 21.5 °C to 41.5 °C and recording the corresponding coincidence, the temporal center position of the coincidence as a function of the working temperature of the MgO: PPLN-WG was also investigated and displayed in Fig. 3(b) by blue dots. Through linear fitting, as given by the red solid line, a temperature-dependent slope of $193.84 \pm 0.46$ ps/°C was obtained. Using the above obtained GDD value to do the numerical transformation, the temperature-dependent frequency tuning coefficient ($\omega_i'$) was pre-calibrated. Upon conversion to the wavelength domain, this coefficient was determined to be $0.58 \pm 0.04$ nm/°C.

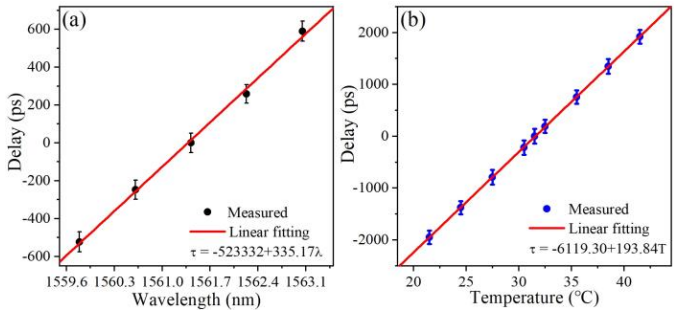

Fig. 3. Measured results of (a) the center time delay variation as a function of the wavelength of the OF, (b) the center time delay variation versus the working temperature of MgO: PPLN-WG. The corresponding linear fittings are given in red lines.

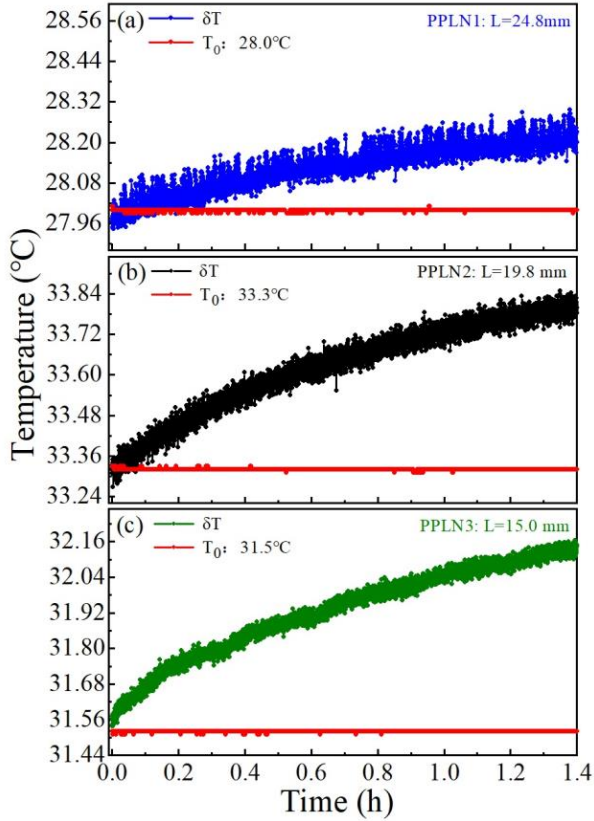

Fig. 4. Measured real-time temperature variations of three MgO: PPLN-WGs under a static TEC-based temperature control using DFT technology. The red dotted lines represent the measured thermistor temperatures for the three WGs when they were set at their optimal phase matching temperatures ($T_0$). The blue, black, and green dotted lines represent the measured temperature variation ($\delta T$) experienced by them, respectively: (a) PPLN1 with a length of 24.8 mm @ 28.0 °C; (b) PPLN2 with a length of 19.8 mm @ 33.3 °C; (c) PPLN3 with a length of 15.0 mm @ 31.5 °C.

Prior to implementing dynamic temperature compensation on the MgO: PPLN-WG, the long-term temperature variation characteristics were evaluated. Three MgO: PPLN-WGs with different lengths were under test, with their accompanied TECs being controlled at a static temperature that satisfies the optimal phase-matching conditions for degenerate down-conversion. The measured results are shown in Fig. 4. To be precise, the three MgO: PPLN-WGs were heated from 25.0 °C to their individual degenerate phase-matching temperature (28.0 °C, 33.3 °C and 31.5 °C, respectively), and the corresponding TEC temperature variation are respectively shown by the red dotted line. Simultaneously, the real-time thermal gradient-induced temperature variations ($\delta T$) were recorded using the aforementioned DFT technique and are depicted in Fig. 4 by distinct dotted lines in blue, black, and green. It is evident that the TEC controller has sufficient stability to maintain a temperature uncertainty within ± 10 mK, which corresponds to a trivial wavelength uncertainty less than ± 6 pm. However, the thermal gradient-induced temperature variations in the three WGs all exhibit significant drift larger than 0.3 °C within the measurement period approximately 1.4 hours. Among them, the largest temperature drift of up to 0.59 °C was observed for the PPLN3-WG, as indicated by the green dotted line in Fig. 4(c). Based on the calibrated temperature-dependent frequency tuning coefficient mentioned above, this temperature drift can result in a significant wavelength drift of up to 342 pm.

Subsequently, the dynamic temperature compensation based on the digital PID algorithm[28] was implemented onto the PPLN3-WG. To evaluate the PID feedback performance, long-term measurements of the idler photons' wavelength at a fixed TEC-controlled temperature of 31.5 °C were conducted. As depicted in Fig. 5(a), when the dynamic temperature compensation is inactive, the center wavelength experiences a drift of more than 180.79 pm (peak-to-peak) along the black line over a period of 14 hours. This drift exhibits an irregular spread with a standard deviation (SD) of 39.08 pm, as indicated by the histogram on the right side. The significant wavelength drift is observed within the initial 7 hours of the measurement, afterward the drift slows down. This phenomenon reveals that realizing the thermal equilibrium inside the WG is an extremely tardy process. In contrast, when adopting the dynamic temperature compensation method, the wavelength drift can be greatly suppressed. As shown by the red line in Fig. 5(b), there is no apparent drift observed in the idler photons' wavelength and the peak-to-peak variation was reduced to ± 23.65 pm. By looking at the variation spread as shown by the histogram on the right side, a normal distribution with a standard deviation of 10.07 pm was observed.

The wavelength stability in terms of ADEV was further measured and shown in Fig. 5(c). One can see that, without applying the dynamic temperature compensation, the ADEV curve (black squares) follows a descending slope of $1/\sqrt{\tau}$ until the averaging time reaches 100 s. Afterward, the ADEV slope gradually flattens out, leading to a degraded stability of $1.41 \times 10^{-5}$ at the averaging time of 10000 s. With the dynamic temperature compensation in operation, the ADEV result is given by red circles. Although a modest degradation of the stability happened at the short-term averaging time within 10 s due to the temperature feedback, a remarkable improvement of the long-term stability was achieved. At the averaging time of 10000 s, the ADEV reached $2.00 \times 10^{-7}$, which is improved by nearly two orders of magnitude.

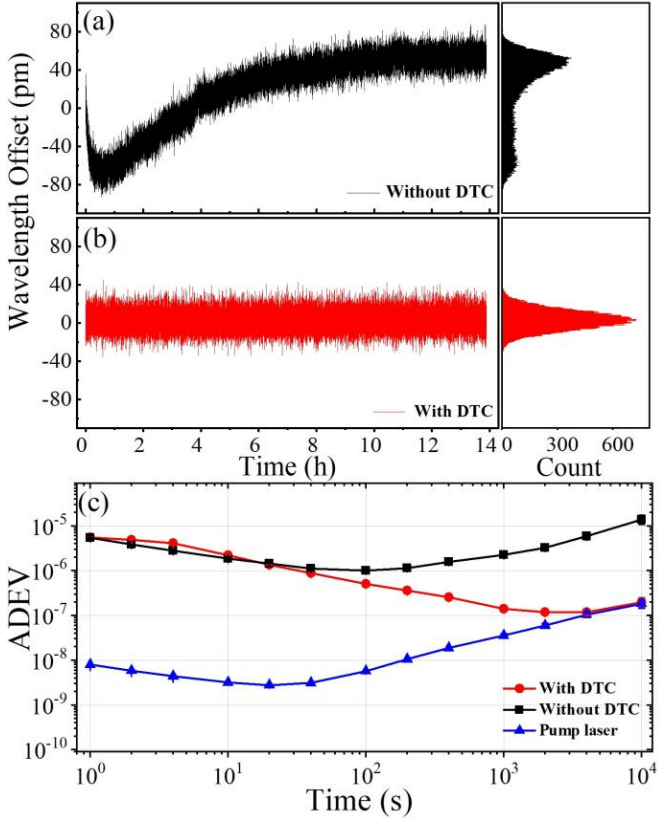

Fig. 5. Wavelength variation of the idler photons during a long-term measurement of lasting for 14h under the condition of without (a) and with (b) the dynamic temperature compensation (DTC). (c) The wavelength stability of the idler photons in terms of Allan deviation (ADEV) for the cases of without (black squares) and with (red circles) the dynamic temperature compensation. The wavelength stability of the pump laser is also given by blue triangles for comparison.

Besides, the long-term wavelength stability of the pump laser was monitored and analyzed. Its ADEV curve is presented in Fig. 5(c) by blue triangles, serving as an indicator of the system's noise floor. From the concurrence between the red dots and blue triangles beyond the averaging time of 5000 s, one can conclude that, the achievable long-term wavelength stability for the idler photons is fundamentally limited by that of the pump.

## Conclusion

In conclusion, we have demonstrated a dynamic temperature compensation method that efficiently stabilizes the wavelength of the entangled biphoton source, generated from a MgO: PPLN-WG. By leveraging the DFT technique, we are able to promptly extract spectral information of the generated photons from the TCSPC histogram. With this information, the thermal gradient-induced temperature variation can be identified and further compensated. Consequently, the long-term wavelength drift of the down-converted photons is significantly suppressed by nearly two orders of magnitude in terms of the Allan deviation, achieving a remarkable value of $2.00 \times 10^{-7}$ @ 10000 s. This result is found primarily constrained by the wavelength stability of the pump laser. In the future, the wavelength stability of the biphoton source can be potentially enhanced by incorporating laser frequency stabilization technique[29] into the pump. The proposed method offers a straightforward and efficient means for measuring and stabilizing the photon wavelength, promising seamless compatibility with off-the-shelf entangled biphoton generation and measurement setups. This advancement is anticipated to expedite the development and application of wavelength-stable entangled biphotons.

**Funding Sources.** This work was supported by National Natural Science Foundation of China (12033007, 61801458, 12103058, 12203058, 12074309, 61875205), the Youth Innovation Promotion Association, CAS (Grant No. 2021408，2022413, 2023425), the China Postdoctoral Science Foundation (2022M723174).